\documentclass[sigconf]{acmart}
\usepackage{colortbl}
\usepackage{color}
\usepackage{multirow}
\usepackage{graphicx}
\AtBeginDocument{%
  \providecommand\BibTeX{{%
    \normalfont B\kern-0.5em{\scshape i\kern-0.25em b}\kern-0.8em\TeX}}}

\copyrightyear{2021}
\acmYear{2021}
\setcopyright{acmcopyright}
\acmConference[ICDAR '21]{Proceedings of the 2021 Workshop on Intelligent Cross-Data Analysis and Retrieval}{August 21--24, 2021}{Taipei, Taiwan}
\acmBooktitle{Proceedings of the 2021 Workshop on Intelligent Cross-Data Analysis and Retrieval (ICDAR '21), August 21--24, 2021, Taipei, Taiwan}
\acmPrice{15.00}
\acmDOI{10.1145/3463944.3469270}
\acmISBN{978-1-4503-8529-9/21/08}

\begin{document}
\title{Two-Faced Humans on Twitter and Facebook: \\ Harvesting Social Multimedia for Human Personality Profiling}

\author{Qi Yang}
\email{yangqi@itmo.ru}
\affiliation{%
  \institution{ITMO University}
  \country{Russia}
  }

\author{Aleksandr Farseev}
\email{farseev@gmail.com}
\affiliation{%
  \institution{ITMO University, SUSS School of Business, SoMin.ai Research}
  \country{Russia, Singapore}
}

\author{Andrey Filchenkov}
\email{afilchenkov@itmo.ru}
\affiliation{%
  \institution{ITMO University}
  \country{Russia}
}  
\begin{abstract}
  Human personality traits are the key drivers behind our decision-making, influencing our life path on a daily basis. Inference of personality traits, such as Myers-Briggs Personality Type, as well as an understanding of dependencies between personality traits and users' behavior on various social media platforms is of crucial importance to modern research and industry applications. The emergence of diverse and cross-purpose social media avenues makes it possible to perform user personality profiling automatically and efficiently based on data represented across multiple data modalities. However, the research efforts on personality profiling from multi-source multi-modal social media data are relatively sparse, and the level of impact of different social network data on machine learning performance has yet to be comprehensively evaluated. Furthermore, there is not such dataset in the research community to benchmark. This study is one of the first attempts towards bridging such an important research gap. Specifically, in this work, we infer the Myers-Briggs Personality Type indicators, by applying a novel multi-view fusion framework, called "PERS" and comparing the performance results not just across data modalities but also with respect to different social network data sources. Our experimental results demonstrate the PERS's ability to learn from multi-view data for personality profiling by efficiently leveraging on the significantly different data arriving from diverse social multimedia sources. We have also found that the selection of a machine learning approach is of crucial importance when choosing social network data sources and that people tend to reveal multiple facets of their personality in different social media avenues. Our released social multimedia dataset facilitates future research on this direction.
\end{abstract}

\maketitle

\section{introduction}
During the past decade, an increasing number of social media platforms have been rapidly emerging and therefore such platforms start playing a vital role in facilitating human interactions worldwide. Since 2012, the average daily social media screen time has increased from 60 minutes to 144 minutes~\cite{minutes}. Furthermore, it has spiked even higher since the start of the COVID-19 disease outbreak~\cite{farseev2020understanding}, when people have been locked at home with the only remaining option of engaging their friends through Social Media.


To maintain high user engagement rates, it is essential for social network conglomerates to position and recommend relevant content according to user interests and online behaviours. For example, extroverted people are more likely to use social media in general as they tend to reveal themselves as enthusiastic, interactive, and therefore forming more social circles around themselves~\cite{difference}. However,  contrarily, introverts were found to be spending significantly more time evaluating the value of each online service they use before a deeper user-service interaction may occur~\cite{LU2010150}.

With such large and diverse data available nowadays on Social Media, it is getting practically impossible to manually distinguish social media users when attempting to provide them with more personalized online experiences~\cite{farseev2018somin}. And therefore, an automated approach to human behaviour pattern understanding on social media is well demanded~\cite{somin}. Unfortunately, nowadays personality profiling still heavily depends on manual procedures like questionnaires and quizzes~\cite{questionnaires2}, and therefore its cost remain unacceptably high limiting its usage in real-time online services, such as social networking websites~\cite{farseev2016bbridge}.

However, automatic personality inference is also known to be a hard task~\cite{twitterfacebook,buraya2018multi}, which is mainly due to the multi-facet nature of social media data. For example, Twitter is often used for casual daily interactions, while Facebook nowadays more perceived as a private communication channel. As a result, Facebook's audiences demographics vary drastically from young to senior ages, while e.g. TikTok's audiences mostly consist of young individuals aged between 18 to 34 years old. Finally, such social networks like Pinterest might not just have a significant audience age shift but also tend to be largely populated by female users~\cite{pinterest}. Furthermore, one might also explore the drastic difference in behavioural traits that people exhibit across various social media avenues. For example, being one of the most open social media outlets, Twitter is known to concentrate on users' expressions rather than their identity, encapsulating our ``real me'' from the broader public~\cite{twitterfacebook}. At the same time, on the specialized personality-focused forums, such as PersonalityCafe\footnote{https://www.personalitycafe.com/}, the communication might be more concentrated on the members' behavioural habits, allowing for gaining a deeper insight into one's behaviour from the content they post. Considering such a multi-facet multi-source cross-demographic environment, the task of automatic personality profiling from social media data appears to be challenging and, being not widely tackled yet by the research community, requires a more in-depth analysis to be accomplished.

Despite the advantages of leveraging multiple data modalities and sources, there are several associated difficulties identified: 
\newline\textbf{Data gathering}. Data from modern social media platforms are often distributed across various Web resources and shielded behind privacy settings. It is therefore important to implement large-scale cross-source data collection techniques.
\newline\textbf{Data representation}. As real-world social media data comes with different data modalities (e.g. text, image, video, location, etc.), the incorporation of such heterogeneous multi-modal data involves the creation of accurate and mutually compatible approaches to data representation (feature learning).
\newline\textbf{Data modelling}: Effective data integration into a single machine learning model is a challenging task, as the data sources and data modalities often represent various aspects of human life and therefore often very different in nature. Even worth, the high dimensionality of the multi-modal feature spaces might often lead to the so-called "curse of dimensionality" problem when being processed directly, and therefore a dimensionality balancing needs to be accomplished.

Inspired by the research gap and challenges above, in this work we raise the following three research questions. First, to establish a benchmark for multi-view personality profiling, it is important to understand: \textbf{(RQ1): Is it possible to reliably and accurately infer user personality traits at a large scale in an automatic fashion?} Second, to gain an understanding of the real-world applicability of our approach to modern social media scenarios, it is crucial to discover if: \textbf{(RQ2) Is it possible to improve personality profiling performance by leveraging multi-view social multimedia data?} Third, to establish a clear path of future research on multi-source learning, it is crucial to know: \textbf{(RQ3) What is the impact of social media data origin on personality user profiling performance?}

To answer our proposed research questions, in this study we introduce a novel multi-view personality profiling meta ensemble framework, called "PERS", which is able to effectively profile social media user personality by leveraging multimodal multimedia data coming from multi-facet social networks. Furthermore, we introduce efficient data gathering and representation techniques, allowing for seamless processing of the data from Facebook, Twitter, and PersonalityCafe social media forums. Finally, we release the PERS dataset\footnote{PERS Multi-Source Multi-View Personality Dataset: https://pers.azurewebsites.net} to the research community, allowing for future extensive cross-disciplinary research. 

The major contributions of this work are threefold. First and foremost, we have proposed a novel machine learning framework for multi-view user profiling and demonstrated that efficient personality profiling is possible and able to achieve industry-level performance for several personality attributes. Second, we have demonstrated that different social networks are different in nature, which impacts the personality profiling performance and therefore needs to be considered during the data modelling process. Third, we have released a new multi-source cross-social personality profiling dataset to be used by the research community in future studies along the direction.

\vspace{-1mm}
\section{Related works}
In the past two decades, there have been several studies conducted, which were attempting to model human personality traits from a statistical perspective. First, being inferred from statistical analysis of the English lexicon, the Big Five model have been proposed by Digman~\cite{Big5}, where the author reveals the close relationship between human personality and their written language. Inspired by the idea, later on, Pennebaker et al.~\cite{Pennebaker1999LinguisticSL} laid the foundation of statistical personality profiling by introducing the LIWC word categorization scheme, which has numerically bridged the personality traits and the written language utilization patterns.

Furthermore, several studies have been devoted to automatic personality profiling, where cross-disciplinary research groups were utilizing machine learning techniques for automatic human personality inference based on test-generated data~\cite{Mairesse, Argamon05}. Worth noting that these studies were all based on relatively small datasets and therefore are limited in supporting large scale observations staying far apart from being applied in a real-world scenario. Moving forward, the ''Small Data'' problem was partially mitigated by the introduction of the ''MyPersonality'' project~\cite{Mypersonality} - the first large-scale personality-labelled dataset that includes user-generated data from Facebook, which has immediately attracted multimedia community attention entailing first larger-scale studies along social media personality profiling research direction~\cite{attention1,attention2,attention3}. 

These studies have made a giant leap in the field, however, one could also notice that most of them still lack a very important factor limiting their real-world applicability - they are largely focused on a single data sources (e.g. Facebook) or a single data modality (i.e. Text), which brings them apart of being applied to modern multi-source multi-view social multimedia data. Namely, the Linguistic Inquiry and Word Count (LIWC) works~\cite{liwc1,liwc2} are mostly focused on text-only data processing to predict personality by using personality-labelled word categories. At the same time, Arnoux et al.~\cite{glove1} and Tandera et al.~\cite{glove2} instead utilized pre-trained word Global Vectors for Word Representation (GloVe) embedding of textual data, first in the world reporting the results of machine-learning-driven single-modal personality inference.

Finally, there were several studies conducted approaching user profiling from a multi-modal data perspective. For example, Farseev et al.~\cite{farseev2015harvesting} have proposed a multi-modal ensemble model tackling the task of demographic profiling from multi-modal data. Furthermore, authors have extended the framework to leverage sensor data and multi-source multi-task learning for wellness profiling~\cite{farseev2017tweetCanBeFit}. Buraya et al.~\cite{buraya2017towards} proposed to solve the problem of relationship status inference via applying ''out of the box'' machine learning on early-fused data from Twitter, Instagram, Facebook, and Foursquare, which has achieved a significant $17$\% inference performance uplift, as compared to single-modal learning. Going further, \cite{Factoriz} proposed a factorization method to model the intra-modal and inter-modal relationships within multi-modal data inputs, which proved the crucial role of multi-modal data incorporation in improving the user profiling performance; while Buraya et al.~\cite{buraya2018multi} instead leveraged on the temporal component of the multi-modal data, being first applying deep learning for multi-view personality profiling. Being a significant contribution to the field of multi-view profiling, the above works still lack multi-source cross-social network data processing~\cite{farseev2017360}, which limits their applicability in the majority of real-world scenarios.

As it can be seen, there is a significant evidence that incorporation of multi-modal data for automatic user profiling is useful for the achievement of a better prediction performance. However, when it comes to evaluating the role of social network choice for user profile learning, the existing research results remain to be relatively sparse. At the same time, it is reasonable to assume that often serving different needs of an individual, various social media sources might provide a very diverse data in nature, and therefore a more comprehensive study on the roles of different data sources for personality user profiling is necessary.

\section{ITMO Multi-Source Multi-View Personality Dataset}
\begin{figure*}
	\centering
	{
	\includegraphics[width=\textwidth]{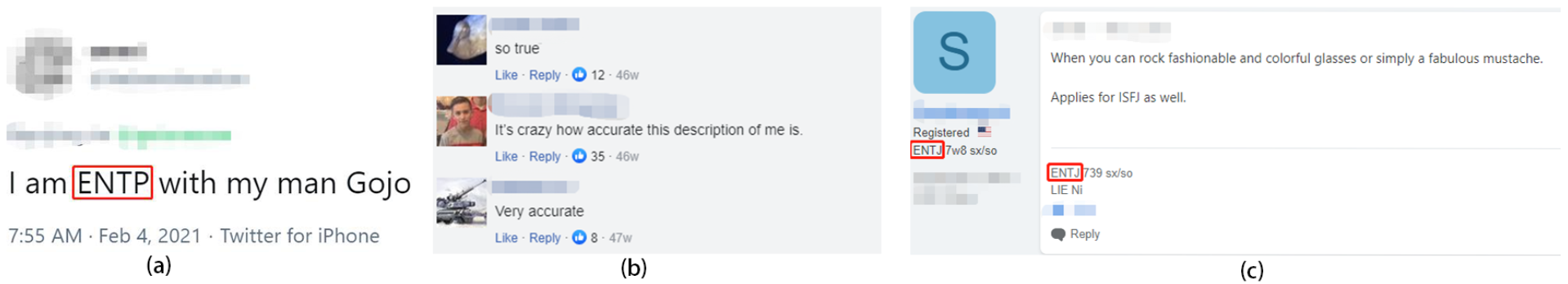}
	\caption{Target user from Twitter (a), Facebook (b), PersonalityCafe (c).}
	\label{fig:example}
	}
	\vspace{-1em}
\end{figure*}

\begin{table}[]
\vspace{-1em}
\caption{Dataset statistics}
\vspace{-1em}
\begin{tabular}{cccc}
\hline
              & Twitter & Facebook & PerCafe      \\ \hline
\#User         & 21305   & 11730 & 3800  \\ 
\#Posts        & 8114568 & 2838141 & 621482\\ 
\#Images       & 1865562 & 597164  & -\\ 
\#Extroversion & 5013    & 6243 & 981    \\ 
\#Introversion & 16292   & 5487 & 2819    \\ 
\#Sensing      & 2799    & 2300 & 610    \\ 
\#Intuition    & 18506   & 9430 & 3190   \\ 
\#Thinking     & 6743    & 3040 & 1666    \\ 
\#Feeling      & 14562   & 8690 & 2134    \\ 
\#Judging      & 9800    & 5528 & 1613    \\ 
\#Perceiving   & 11505   & 6202 &2187   \\ \hline
\end{tabular}

\label{statistics}
\end{table}

\begin{table}[]

\caption{Proportion of 4 personality categories in Twitter, Facebook and PersonalityCafe datasets.}
\vspace{-1em}
\begin{tabular}{cccc}
\hline
               & Twitter & Facebook      & PerCafe \\ \hline
\#Extroversion & 23.53\% & 53.22\% & 25.82\% \\
\#Introversion & 76.47\% & 46.78\% & 74.18\% \\
\#Sensing      & 13.14\% & 19.61\% & 16.05\% \\
\#Intuition    & 86.86\% & 80.39\% & 83.95\% \\
\#Thinking     & 31.65\% & 25.92\% & 43.84\% \\
\#Feeling      & 68.35\% & 74.08\% & 56.16\% \\
\#Judging      & 46.00\% & 47.13\% & 42.45\% \\
\#Perceiving   & 54.00\% & 52.87\% & 57.55\% \\ \hline
\end{tabular}
\vspace{-1em}
\label{percentage}
\end{table}

\begin{table}[]
\vspace{-1em}
\caption{Personality traits distribution}
\vspace{-1em}
\begin{tabular}{cccc}
\hline
     & \textbf{PerCafe} & \textbf{Twitter} & \textbf{Facebook} \\ \hline
INFP & 713     & 5334    & 1665              \\
INFJ & 664     & 4177    & 1498              \\
INTP & 508     & 1121    & 814               \\
INTJ & 487     & 3544    & 521               \\
ENFP & 353     & 3496    & 2381              \\
ENTP & 256     & 122     & 671               \\
ISFP & 137     & 413     & 161               \\
ISTP & 127     & 508     & 131               \\
ENTJ & 113     & 389     & 412               \\
ISTJ & 98      & 739     & 162               \\
ENFJ & 96      & 323     & 1468              \\
ISFJ & 85      & 456     & 535               \\
ESTP & 50      & 200     & 63                \\
ESFJ & 43      & 52      & 666               \\
ESFP & 43      & 311     & 316               \\
ESTJ & 27      & 120     & 266               \\ \hline
\end{tabular}\label{userdistribution}
\vspace{-2em}
\end{table}

\subsection{Data Acquisition}\label{sec:data}
To represent human personality, in this work we use the Myers-Briggs Type Indicator (MBTI)~\cite{myers1998mbti}, which has been widely adopted by the research community~\cite{buraya2018multi,buraya2017towards} and splits one's personality into 16 types, each formed by the following four binary dimensions:
\begin{itemize}
     \item \textbf{E}xtroversion and \textbf{I}ntroversion (EI): this dimension determines how an individual focuses her energies and interest, whether she is influenced externally by the opinion and interpretation of others (Extroverts) or motivated by her inner thoughts (Introverts).
    \item \textbf{S}ensing and i\textbf{N}tuition (SN): this aspect demonstrates how people interpret knowledge. Sensing personalities make decisions based on their five senses and solid observation, whereas Intuitive individuals favour imagination to constancy.
    \item \textbf{T}hinking and \textbf{F}eeling (TF): a person with Thinking aspect, always exhibits logical behaviour in their decisions, while Feeling individuals are empathic and give priority to emotions over logic.
    \item \textbf{J}udging and \textbf{P}erceiving (JP): this dichotomy describes an individual approach towards work, decision-making, and planning. Judging individuals are highly organized in their thoughts, while Perceivers behave more spontaneously.
\end{itemize}

The data was collected from Twitter, Facebook, and PersonalityCafe\footnote{https://www.personalitycafe.com/} social networks, during the time interval of $1$st Jan $2018$ to $1$st Jan $2021$, and via the two following steps:

\textbf{1) Ground truth collection.} To obtain personality ground truth from Twitter, we have downloaded all of the tweets which contain self-reported personality-related keywords/phrases such as “I’m an \textbf{ENTP}” or "I am an \textbf{ENTP}" \textbf{and extract the personality trait from those phrases to be the ground truth for each user} (see Figure~~\ref{fig:example}(a) for example). To harvest Facebook ground truth, we have monitored Facebook comments under personality test results released on 16personalities portal (see Figure~\ref{fig:example}(b) for example). Likewise, to obtain the personality-related ground truth from the PersonalityCafe forum, we downloaded user's publicly-available self-reported personality traits on their profile pages (see Figure~\ref{fig:example}(c) for example).

\textbf{2) User-generated content (UGC) collection.} To establish UGC collection from Twitter and Facebook, we have downloaded user timelines through Twitter REST API\footnote{https://developer.twitter.com/} and Facebook GRAPH API\footnote{https://developers.facebook.com/}, respectively. At the same time, to collect UGC from the PersonalityCafe forum, we downloaded posts from the MBTI forum thread.

\textbf{3) Data pre-processing.} Since social network data might exhibit significant noise levels and often contain grammatical errors, it's necessary to perform data prepossessing prior to the data modelling stage. At the same time, it is necessary to remove the direct personality mentions from the text content so that the model might not be able to use personality abbreviations from the post content at the inference stage. To mitigate the above two problems, we have pre-processed our dataset via the following three steps: \textbf{1) Data Filtering} To ensure the sufficient amount of data per user for training and inference, we have filtered out the users with less than 10 tweets available;  \textbf{2) In-line label replacement}  For all personality traits, the personality type name was then replaced with the "<type>" placeholder (e.g. "ENTJ" might have been replaced with ''<type>''); \textbf{3) Social indicator replacement}: similarly to Nguyen et al.~\cite{bertweet}, we have further converted emojis into the corresponding descriptive textual strings, removed all non-ASCII words, and normalized the text by replacing user mentions, URLs, hashtags, date-time by the corresponding placeholders as follows: @USER, HTTPURL, HASHTAG, DATETIME.  

\begin{figure}
    \centering
	\includegraphics[width=0.49\textwidth]{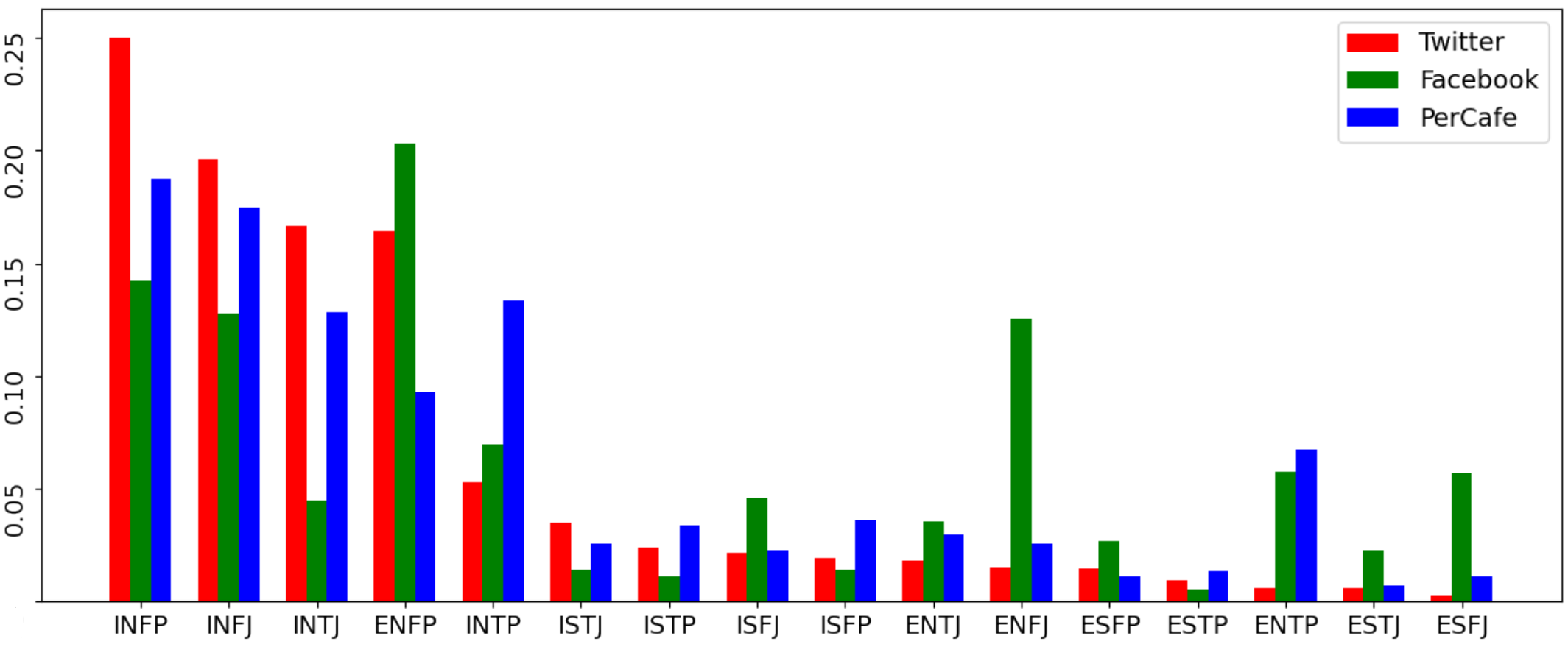}
	\caption{Proportion of personality traits in three data sources}
	\vspace{-1em}
	\label{fig:portion}
\end{figure}

Table~\ref{statistics} highlights the statistics of our dataset across binary personality labels while Figure~\ref{fig:portion} visually reflects the personality label distributions. From the Table~\ref{userdistribution} and the Figure~\ref{fig:portion}, it can be seen that INFP, INFJ, INTJ, ENTP, INTP are the Top-5 most popular personality labels found in both Facebook and Twitter datasets, showing the consistency of the data distributions across general social networks,  reducing the risk of falling into source-dependent bias during the data modeling stage. At the same time, it is important to note that in the PersonalityCafe forum data, the ENFP, INFP, INFJ, ENFJ  labels dominate the rest. The latter observation shows that the personality-related data sources might have a distribution shift towards individuals of certain personality types (different from a general distribution) that tend to participate in such specific personality-related discussions. Therefore, the evaluation based on PersonalityCafe dataset must be accomplished independently.

\subsection{On MBTI Personality Categorization}\label{scale}
The MBTI personality categorization scheme defines each of the $4$ binary MBTI categories to represent a different aspect of human personality. However, when being combined into $16$ personality types, it is known to be associated with a major shortcoming of the overlap between the ''neighbour categories'' (e.g. INTJ and INTP). Given the noisy nature of the content from Social Media, it might be a good idea to predict individual binary MBTI personality traits, instead of modelling the overlapping 16-category scenario. Therefore, in this work, we have adopted such binary personality categorization scheme.

\vspace{-1mm}
\subsection{Data Representation}\label{representation}
To facilitate an effective data modelling process, the data needs to be properly represented in the form of feature vectors. Following the best practices described in the literature on user profiling~\cite{Khan2020}~\cite{panRecent}~\cite{farseev2017tweetFit} we have chosen the following data representation approaches:

\textbf{1) Textual Features:} First, to represent the textual data at a user level, for each user all posts were concatenated into a corresponding user-specific ''documents''. Second, the term frequency-inverse document frequency (TF-IDF) has been extracted to form the document-term matrix. Finally, we have applied the Latent Semantic Analysis (LSA~\cite{LSA}), as such transformation has previously shown sizable performance uplift~\cite{Daneshvar2018} when being applied for user profiling. The final dimension of the compressed textual feature vector was set to $100$, where the new number of dimensions has been found empirically during a grid search.

\textbf{2) Visual Features:} To represent visual data, we have automatically mapped each photo into the distribution of $1000$ ImageNet~\cite{imagenet} image concepts via the pre-trained ResNet-101 model~\cite{Resnet}. We then summed up the predicted concept occurrence likelihoods for each user and element-vice normalized the obtained vector by the total number of images available from the user. In such a way, for each user, we have obtained a $1000$-sized image concept distribution vector. Similarly to the text modality, Principal Component Analysis (PCA~\cite{Jolliffe2011}) has been further applied to reduce the dimensionality of the visual feature space to $200$.

\section{Problem definition}
\subsection{Social media data}\label{def}
Given a user \textit{i}, we denote the multi-view data associated with the user as a set \textit{X}:
\begin{equation}
    X= \{(y_i,  T_i,   I_i), i=1,2,\ldots,l \}
\end{equation}
where $l$ represents the number of samples in the dataset, $T_I$ is the collection of the $i$th user text content represented as Textual Features, $I_i$ is the collection of $i$th user image content represented as Image features, and $y_i$ represents one of the four $i$th binary personality trait ground truth labels.

In such a way, we can now formulate the personality profiling as a user-level multi-view document classification task, where users play the role of ''documents'' when being rephrased in standard terms. 

\begin{figure*}
    \centering
	\includegraphics[width=0.85\textwidth]{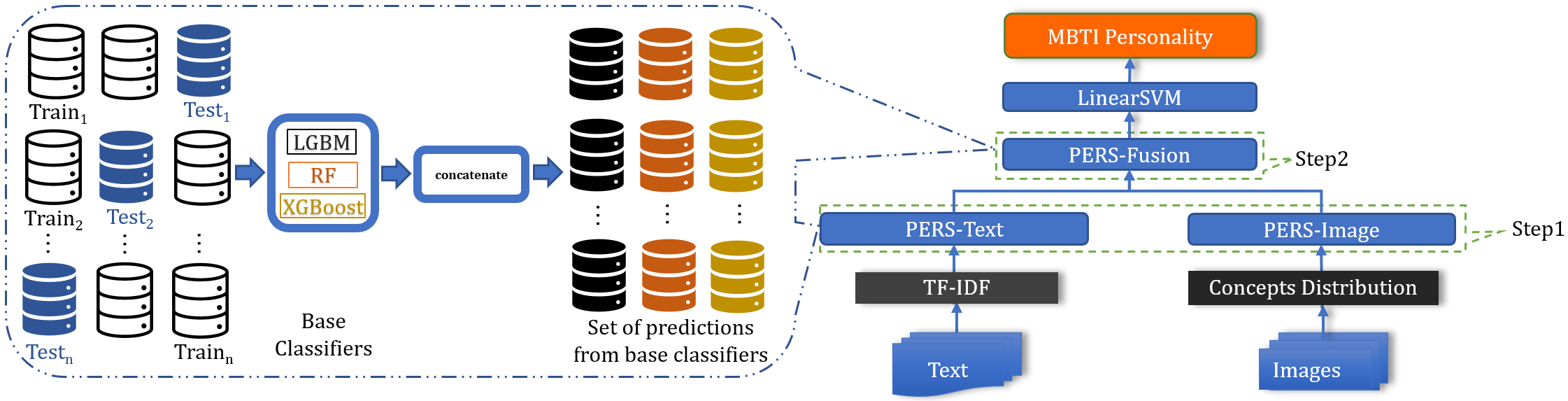}
    \caption{Overview of the PERS framework}
    \vspace{-1em}
	\label{fig:Model}
	
\end{figure*}

\subsection{PERS Framework}\label{pers}
Below, we now can define the PERS framework as a two-step stacked generalized ensemble approach. The architecture of our PERS framework is illustrated in Figure~\ref{fig:Model}.

\textbf{First-Step}: 
Given the training set $X= \{(y_i, x_i), i=1,2,\ldots,I \}$ , where $y_{i}$ is the personality label and $x_{i}$ represents the feature vector, we shuffle and uniformly split $X$ into $K$ equal sub-sets. In such a way, $X_{k}$ and $X^{(-k)} = X - X_{k}$ are the definitions of the test and training sets, respectively, for the $k$-th fold in K-Fold cross-validation.

Now, we can specify a J-sized list of ''base'' classifiers and train the $j$-th base classifier based on the training set $X^{(-k)}$. We denote the $Z_{ki}$ as the prediction of $j$-th classifier on $x_{i}$ for each sample in the test set for the $k$-th cross-validation fold. In such a way, immediately after the cross-validation process, we can define a new dataset based on J ''base'' classifier outputs:
\begin{equation}
    Z_{first} = \{(y_i,z_{1i},\ldots,z_{Ji}),i=1,\ldots,I\}
\end{equation}

For each data modality, we compute $Z_{Tfirst}$ and $Z_{Ifirst}$ separately and then we form the input for the next stage of data processing $H$ via column-wise concatenation of $Z_{Tfirst}$ and $Z_{Ifirst}$. The corresponding dimension of $H$ is therefore $n\times 2J$.

\textbf{Second-Step}: 
Same as the first step, we perform K-fold cross-validation on each of $j$ base classifier results with $H$ coming from First-Step to obtain the output $Z_{second}$ formally defined as follows:

\begin{equation}
Z_{second} = \{(y_i,w_{1i},\ldots,w_{Ji}),n=1,\ldots,I\}
\end{equation}
where, $w_{Kn}$ represent the $i$-th prediction inferred by the $k$-th base classifier. The corresponding dimension of $Z_{second}$ is therefore $n\times J$.

Therefore, to get the final model prediction, we trained a meta classifier $f$ on $Z_{second}$, which can be formulated as :
\begin{equation}
    y = f(Z_{second}).
\end{equation}
Specifically in this study, we have chosen Support Vector Machine with linear kernel(LinearSVM) as our meta classifier.

\subsection{Base Classifiers}\label{baseclassifier}
 To maximize the performance of the PERS framework, it is of a crucial importance to choose a set of suitable machine learning algorithms as base classifiers. The previous studies ~\cite{mti4010009,Qi_Aleksandr_Andrey_2020,farseev2015harvesting} suggest XGBoost~\cite{XGboost}, LightGBM~\cite{LGBM}, and Random Forest to be the top-choice base models for user profiling, as their performance on social media data has been reported to beat state-of-the-art baselines, often including those baselines coming from Deep Learning community. 
 
 \subsubsection{Random Forest}
Random forest is an ensemble learning algorithm that integrates multiple decision trees to complete prediction. For the classification problem, the prediction result is the vote of all the decision tree prediction results. During training, bootstrap sampling is used to form the training set of each decision tree. When training each node of each decision tree, the features used are also part of the features extracted from the entire feature vector. By integrating multiple decision trees, and training each decision tree with sampled samples and feature components each time, the variance of the model can be effectively reduced. 
 
 \subsubsection{XGBoost}
XGBoost is an effective and scalable gradient booster machine that has been widely adopted in the industry domain in recent years. It is an ensemble model containing a set of classification and regression trees (CART). Given a training data $x_{i}$ and target $y_{i}$, the XGBoost model can be defined as:
 \begin{equation}
     \hat{y}_i=\sum_{K=1}^{K} f_k(x_i), f_k\in F
 \end{equation}
where K is the total number of trees, $f_k$ for $k_th$ tree is a function in the functional space F, and F is the set of all possible CARTs.

 \subsubsection{LightGBM}
LightGBM is the improved version of Gradient Boosting Machines that mitigates the ''optimal division point search'' problem, arising from the increasing computational complexity on larger datasets. The problem is solved via the following two tricks, reducing the training data size and data dimensionality:
  
\textit{Gradient-based One-Side Sampling (GOSS)}: Exclude most of the samples with small gradients, and only use the remaining samples to calculate the information gain. 

\textit{Exclusive Feature Bundling (EFB)}: Bundles mutually exclusive features as they rarely take non-zero values at the same time.

\section{Evaluation}
To answer our research questions, we have evaluated the performance of PERS framework being trained across different data sources and data modality combinations. For all experiments, the dataset was uniformly split into training set and test set with the ratio of 85:15, maintaining the original personality label distributions. To understand the impact of different modalities, data sources, and fusion strategies on the final performance of the model, we have selected the following community-adopted personality profiling baselines~\cite{buraya2017towards, farseev2015harvesting, rangel2015overview}:

\begin{itemize}
    \item \textbf{Independently-trained base classifiers} (see descriptions in Section~\ref{baseclassifier}) with respect to each data modality;
    \item \textbf{Early fusion}. Base classifiers trained based on the early-fused data modality representations (concatenated feature vectors);
    \item \textbf{Early Fusion (PCA 200)}. Base classifiers trained based on the early-fused data modality representations with PCA being applied after the vector concatenation (the PCA dimension of $200$ has been selected empirically via a grid search);
\end{itemize}

\subsection{Evaluation Metrics}
Due to the imbalanced distribution of personality labels in our Datasets (see Section~\ref{sec:data} for details on the Data Distributions), for performance evaluation, we have adopted the "\textit{$F_{1macro}$}" metric~\cite{farseev2015harvesting}, which is the harmonic mean between precision and recall, and the average is calculated per label across all labels. The \textit{$F_{1macro}$} metric is formally defined as:
\begin{equation}
    F_{1macro}= \frac{1}{Q}\sum_{j=1}^{Q}\frac{2\times\mathit{pj}\times\mathit{rj}}{\mathit{pj}+\mathit{rj}}
\end{equation}
where $p_{j}$ and $r_{j}$ are the precision and recall for $\lambda_{j}$ $\in$ h($x_{i}$) from $\lambda_{j}$ $\in$  $y_{i}$.

We have further adopted the ~\textit{Matthews correlation coefficient} metric ~\cite{mcor} (Mcor), as it incorporates both true and false positives and negatives and generally regarded as a ''balancing'' measure that can be used even if the classes are of a very different size. The \textit{Mcor} metric is formally defined as:
\begin{equation}
    Mcor = \frac{TP * TN - FP * FN}{\sqrt{(TP + FP) * (FN + TN) * (FP + TN) * (TP + FN)}}
\end{equation}
where TP is the number of true positives, TN the number of true negatives, FP the number of false positives and FN the number of false negatives.

We prioritize the \textit{$F_{1macro}$} score as our main evaluation metric, while the \textit{Mcor} score plays an auxiliary role for making decisions regarding performance when the \textit{$F_{1macro}$} values are marginal.

\begin{table*}[]
\caption{Evaluation of the "PERS" framework trained on the independent modality and the modality combinations in Twitter and Facebook. Text in green indicate the best performance while red indicate the worst.}
\begin{tabular}{ccccccccc}
\hline
\multicolumn{1}{c|}{\multirow{2}{*}{\textbf{Model}}} & \multicolumn{4}{c|}{\textbf{Twitter}} &\multicolumn{4}{c}{\textbf{Facebook}} \\ \cline{2-9} 
\multicolumn{1}{c|}{} &EI  &SN  &TF &\multicolumn{1}{c|}{JP}&EI &SN &TF &JP \\ \hline
\multicolumn{9}{c}{Text($F_{1macro}$/Mcor)} \\ \hline
\multicolumn{1}{c|}{XGBoost}                          & 0.80/0.62                         & 0.48/0.08                         & 0.59/0.23                         & \multicolumn{1}{c|}{0.61/0.22}                         & 0.62/0.25                         & 0.59/0.22                         & 0.55/0.16                         & 0.58/0.17                         \\
\multicolumn{1}{c|}{RF}                               & 0.78/0.56                         & 0.56/0.14                         & 0.61/0.22                         & \multicolumn{1}{c|}{0.62/0.25}                         & 0.61/0.21                         & 0.62/0.26                         & 0.58/0.17                         & 0.58/0.16                         \\
\multicolumn{1}{c|}{LGBM}                             & 0.80/0.62                         & 0.56/0.14                         & 0.61/0.24                         & \multicolumn{1}{c|}{0.62/0.25}                         & 0.62/0.24                         & 0.62/0.24                         & 0.57/0.13                         & 0.58/0.16                         \\ \hline
\multicolumn{9}{c}{Image($F_{1macro}$/Mcor)}                                                                                                                                                                                                                                                                                                                                         \\ \hline
\multicolumn{1}{c|}{XGBoost}                          & \textcolor{red}{0.46/0.01} & \textcolor{red}{0.47/0.03} & \textcolor{red}{0.51/0.01} & \multicolumn{1}{c|}{\textcolor{red}{0.54/0.09}} & 0.59/0.19 & 0.47/0.05 & \textcolor{red}{0.49/0.07} & 0.56/0.14                         \\
\multicolumn{1}{c|}{RF}                               & 0.54/0.09                         & 0.52/0.06                         & 0.57/0.14                         & \multicolumn{1}{c|}{0.56/0.12}                         & 0.59/0.18                         & 0.54/0.08                         & 0.57/0.15                         & 0.57/0.14                         \\
\multicolumn{1}{c|}{LGBM}                             & 0.52/0.05                         & 0.52/0.04                         & 0.56/0.11                         & \multicolumn{1}{c|}{0.57/0.13}                         & 0.59/0.18                         & 0.55/0.11                         & 0.57/0.14                         & 0.56/0.12                         \\ \hline
\multicolumn{9}{c}{Early Fusion($F_{1macro}$/Mcor)}                                                                                                                                                                                                                                                                                                                                  \\ \hline
\multicolumn{1}{c|}{XGBoost}                          & 0.8/0.62                          & 0.48/0.06                         & 0.59/0.22                         & \multicolumn{1}{c|}{0.59/0.19}                         & 0.60/0.21                         & 0.56/0.22                         & 0.54/0.16                         & 0.58/0.17                         \\
\multicolumn{1}{c|}{RF}                               & 0.78/0.56                         & 0.54/0.1                          & 0.62/0.24                         & \multicolumn{1}{c|}{0.62/0.24}                         & 0.64/0.27                         & 0.62/0.24                         & 0.60/0.20                         & 0.57/0.15                         \\
\multicolumn{1}{c|}{LGBM}                             & 0.80/0.61                         & 0.55/0.11                         & 0.62/0.25                         & \multicolumn{1}{c|}{0.62/0.24}                         & 0.62/0.24                         & 0.61/0.22                         & 0.60/0.21                         & 0.60/0.20                         \\ \hline
\multicolumn{9}{c}{Early Fusion(PCA 200)($F_{1macro}$/Mcor)}                                                                                                                                                                                                                                                                                                                         \\ \hline
\multicolumn{1}{c|}{XGBoost}                          & 0.79/0.61                         & 0.48/0.07                         & 0.59/0.23                         & \multicolumn{1}{c|}{0.60/0.20}                         & \textcolor{red}{0.57/0.14} & \textcolor{red}{0.47/0.04} & 0.50/0.07                         & \textcolor{red}{0.55/0.10} \\
\multicolumn{1}{c|}{RF}                               & 0.78/0.56                         & \textcolor{green}{0.57/0.14} & \textcolor{green}{0.63/0.26} & \multicolumn{1}{c|}{0.62/0.24}                         & 0.60/0.21                         & 0.54/0.08                         & 0.58/0.16                         & 0.56/0.13                         \\
\multicolumn{1}{c|}{LGBM}                             & 0.80/0.60                         & 0.56/0.11                         & 0.62/0.25                         & \multicolumn{1}{c|}{0.62/0.24}                         & 0.59/0.18                         & 0.52/0.04                         & 0.56/0.12                         & 0.56/0.13                         \\ \hline
\multicolumn{9}{c}{PERS Trained with Single Modality($F_{1macro}$/Mcor)}                                                                                                                                                                                                                                                                                                                        \\ \hline
\multicolumn{1}{c|}{Text}                             & 0.81/0.62                         & 0.55/0.14                          & 0.62/0.26                         & \multicolumn{1}{c|}{0.63/0.28}                         & 0.62/0.23                         & 0.62/0.28                          & 0.59/0.21                         & 0.58/0.16                         \\
\multicolumn{1}{c|}{Image}                            & 0.53/0.11                         & 0.47/0.05                         & 0.57/0.16                         & \multicolumn{1}{c|}{0.59/0.17}                         & 0.59/0.17                         & 0.50/0.10                         & 0.56/0.17                         & 0.56/0.13                         \\ \hline
\multicolumn{9}{c}{PERS Trained with Dual Modalities($F_{1macro}$/Mcor)}                                                                                                                                                                                                                                                                                                                          \\ \hline
\multicolumn{1}{c|}{T+I}                              & \textcolor{green}{0.82/0.61} & 0.54/0.12                         & \textcolor{green}{0.63/0.26} & \multicolumn{1}{c|}{\textcolor{green}{0.64/0.28}} & \textcolor{green}{0.64/0.28} & \textcolor{green}{0.63/0.30} & \textcolor{green}{0.61/0.23} & \textcolor{green}{0.62/0.21} \\ \hline
\end{tabular}
\label{Twitterfacebook}
\end{table*}

\begin{table}[]
\caption{Evaluation of the "PERS" framework trained on text modality on PersonalityCafe.}
\vspace{-1em}
\begin{tabular}{ccccc}
\hline
\multicolumn{5}{c}{Text($F_{1macro}$/Mcor)}                                                                                 \\ \hline
\multicolumn{1}{c|}{model}    & EI                 & SN                 & TF                 & JP                 \\ \hline
\multicolumn{1}{c|}{LGBM}     & 0.69/0.39          & 0.74/0.50          & 0.80/0.61          & 0.73/0.47          \\
\multicolumn{1}{c|}{XGBoost}  & 0.69/0.41          & 0.69/0.42          & 0.79/0.60          & 0.73/0.46          \\
\multicolumn{1}{c|}{RF}       & 0.65/0.29          & 0.67/0.33          & 0.74/0.53          & 0.69/0.36          \\ \hline
\multicolumn{1}{c|}{PERS} & \textbf{0.71/0.43} & \textbf{0.74/0.51} & \textbf{0.81/0.61} & \textbf{0.74/0.49} \\ \hline
\end{tabular}
\vspace{-1em}
\label{perc}
\end{table}

\subsection{Evaluation Across MBTI Categories}
To judge the applicability of PERS framework in a real-world scenario, we have evaluated the limits of PERS's performance across Twitter, Facebook, and PersonalityCafe datasets. The evaluation results are presented in the Table~\ref{Twitterfacebook}.

From the table, it can be seen that,  being trained on the multi-view data from Twitter, PERS framework is able to achieve an industry-level performance of $0.82$ $F1_{macro}$ score when predicting the Extroversion-Introversion (EI) personality trait. While the performance obtained for the other three personality categories is significantly lower (ranging from -0.18 $F1_{macro}$ score for Judging-Perceiving(JP) to -0.28 $F1_{macro}$ score for Sensing-Intuition(SN)), we believe that such high promising performance for the EI label could testify the tremendous potential of multi-view social media data for psycho-graphic discovery and personality profiling. The superiority for the EI label, at the same time, can be explained by the natural difference of these two human personality categories when it comes to user communication on social platforms: Extroverts are known to be much more open to others, while Introverts - opposite, being more selective and making decisions at a slightly more conservative pace. Such an inspiring results allow us to give a \textbf{positive answer to our RQ1} and possibly gives birth to a wide range of new research directions related to Personality Profiling and Multi-View learning.

But what about other two data sets? An interesting finding comes from the results presented in Table~\ref{perc}, where PERS demonstrates a breakthrough performance based on PersonalityCafe dataset showing the best overall $F1_{macro}$ scores when predicting all $4$ binary MBTI categories. Such a result can be explained by the specific nature of the PersonalityCafe dataset, where users purposely reveal their behavioural differences and therefore often biased towards particular social behaviour concepts. Such results also \textbf{Reassure our positive answer to the Q1} and allow us to conclude that indeed the nature of a data source and the social network use patterns are of the crucial importance when solving the multi-view cross-media personality profiling problem.

\vspace{-2mm}
\subsection{Evaluation Across Different Modalities}
First, we have investigated the contribution of different data modalities towards personality profiling performance and its integration ability. An interesting observation comes from the cross-modal experimental results presented in Table~\ref{Twitterfacebook}: the PERS framework have performed 2\% better than other single-source baselines for all but SN binary labels being trained on Twitter and Facebook datasets.  Another interesting observation can be made from the modality combination results, where being trained based on both textual and image data, PERS is able to outperform by more than 1\% not just other single-modal classifiers but also the early-fused baselines.

The above findings suggest that the introduction of multi-modality into user profiling could serve as a powerful booster of model performance. Such observation could be explained by the richness of visual data when reflecting user preferences, which serves as a greatly beneficial supplement of the textual data modality at the data modeling stage. The latter finding confidently positively answers our \textbf{RQ2} by emphasizing the important role of multi-modal data learning for personality profiling application.

Finally, let us also highlight an interesting observation that comes from single-modal evaluation results (see Table~\ref{Twitterfacebook}). It is important to note that, in the cases of learning from single-modal source, "PERS" being trained on text-modality performs best across all personality labels, ranging from $0.02$ to $0.28$ $F1_{macro}$ score superiority level. The latter can be easily explained by the quantitative domination of textual data over visual modality (see Table~\ref{statistics}).  Another potential reason behind such trend could be the high level of noise in the user-generated visual data, where the images are less strict in terms of perspective and object positioning as compared to professional photos. Moreover, such visual content often includes objects that might not directly reflect the semantics of the data and therefore might be not accurate in representing an author's personality. To this end, such hypothesis also aligns well with our-chosen visual data representation approach, where ImageNet concept distribution might be simply too general for personality profiling tasks, as opposed to, for example, demographic profiling~\cite{farseev2015harvesting}.

\vspace{-1mm}
\subsection{Evaluation Across Different Sources}
At last, let's examine the impact of the social media data origin on personality user profiling performance, so that an industry guideline can be established for future research.

As the textual modality has participated in all three data sources, let's describe the PERS performance on textual data first. From the Table~\ref{Twitterfacebook} and Table~\ref{perc} it can be noticed that PERS framework being trained on Twitter dataset outperformed the performance on Facebook dataset and PerosnalityCafe dataset by more than $0.19$ $F1_{macro}$ score in predicting EI label. In the contrast, when it comes to the SN label, Twitter-trained PERS was not able to outperform Facebook and PersonalityCafe data, staying behind by $0.2$ and $0.11$ $F1_{macro}$ score, respectively. Finally, the PERS performance of TF and JP labels based on Twitter textual data was found to be better than Facebook by $0.03$ and $0.05$ $F1_{macro}$ score, respectively,  but considerably worse than PersonalityCafe by $0.19$ and $0.11$ $F1_{macro}$ score, respectively.

The superiority of Twitter in predicting the EI label could be explained by the differences of the ''energy'' source for Extroverted and Introverted personality types. Precisely, according to Martin~\cite{martin1997looking}, Extroverts prefer to source their life energy from active involvement in events and engaging into different activities, while Introverts often prefer doing things alone obtaining their energy from dealing with the ideas, pictures, memories, and reactions that are inside on their mind. Similarly, from the digital world, it can be seen that on Twitter both personality types are able to express themselves fulfilling both their enjoyment (ENJ) and observation/learning (LEN) needs, while for Facebook ENJ factor got fulfilled proportionally for a smaller number of individuals, affecting the overall user base distribution~\cite{syn2015social}. Correspondingly Twitter and PersonalityCafe data are diverse enough to differentiate the EI personalities achieving a higher prediction scores, as compared to Facebook-based prediction. The observation is also supported by our data distribution (see Section~\ref{sec:data}), where Twitter and PersonalityCafe datasets are clearly skewed towards Introverts, proving more data for PERS to learn on how the personality type direct their energy and make decisions. The latter aspect is important as it is known that Extroverts might generate substantially more UGC as compared to Introverts~\cite{syn2015social} and therefore Introvert-crafted content sufficiency is crucial for mutually-consistent and comprehensive learning from the data.

At the same time, an inverse picture can be noticed for the SN label results, and there is a ''low hanging fruit'' explanation of the phenomena: for both Twitter and Facebook the SN personality is distributed with a clear shift towards Intuitive personality type, while being short on Sensing individuals in the data. Despite reflecting the real life distribution, this data property also entails a possible technical issue of variance insufficiency limiting the model differentiation when it comes to learning the Sensing and Intuitive user personas. Considering that on Facebook and PersonalityCafe there were more Sensing personality types identified, it is reasonable to assume that this is also the reason why PERS have performed better on these latter two sources as compared to the former one. To the end, a more ''sensing'' Facebook can also be explained by the fact that Facebook is mainly treated nowadays as a communication tool so that people land there for fulfilling their daily communication needs, while Twitter more often serves as a source of Inspiration attracting even more Intuitive individuals into its nets~\cite{syn2015social}.

Now, its time to compare the visual modality performance and the first thing that might capture a reader's attention is that the image data modality has performed similarly for the cases of TF and JP labels for both Twitter and Facebook sources, however, at the same time Facebook performed better for EI and SN labels with $0.06$ and $0.03$ $F1_{macro}$ score performance uplift, respectively.  As it has been described earlier, both the personality categories are very different in the way they direct the energy and perceive the external world~\cite{martin1997looking} and therefore the data diversity introduced by incorporation of the visual modality is of crucial importance for the PERS performance. As Twitter is a ''less visual''  data source as compared to Facebook and also its data distributions are less balanced (as discussed above) for both personality labels, it is reasonably to assume that these two factors might entail the superiority of Facebook over Twitter on visual data in our particular case.

Finally, it is worth noting that PERS trained only based on the textual data from PersonalityCafe forum outperforms results obtained from both Twitter and Facebook data by at least 0.1 $F1_{macro}$ score. Such a finding can be easily explained by the precise focus of the PersonalityCafe forum on the personality topic, which provides additional meaningful data descriptors which can be utilized by PERS for improving its personality inference score.

Backed up by all the observations above, we now can give an answer to \textbf{the RQ3} by highlighting the drastic difference of Social Media data sources when being used in automated personality profiling, which is dictated by the way different personalities engaged into social network activities.

\section{Limitations and Future Work}
Although PERS outperforms baselines for all binary personality inference tasks, being combined together, the predicted labels might often mismatch the actual final user MBTI personality score and therefore only binary personality predictions (such as EI prediction for Twitter dataset) can be leveraged in a real world settings. 

Therefore, it is evident that new data source-specific multi-view learning approaches need to be implemented~\cite{farseev2017tweetFit, farseev2017cross} where personality profiling modelling will be leveraging on more multi-view data representations (such as avatar~\cite{gao2013improving}, sensor data~\cite{farseev2017tweet}, etc.) and  mitigating the specific issues arising from the difference of communication styles across different social avenues. The development of such models and their application for content generation or recommendation services will be the focus of our future research.

\section{Conclusions}
In this work, we presented a brave study on automated human personality profiling across multiple data modalities and social networking sites, such as Facebook, Twitter, and PersonalityCafe. Our-proposed personality profiling framework, called "PERS'' have demonstrated an industry-level superior performance over single-source and multi-source baselines. Via our cross-social evaluation, we have also proved that different social networking platforms exhibit various distinct user communication and usage patterns, which in turn affects user profiling model performance and needs to be treated with care for skewed distributed datasets. Finally, to facilitate future research in this exciting direction we have released our new large-scale cross-social multi-view personality profiling dataset and supplemented it with the corresponding statistics and analytics for the community use.

\section{Acknowledgement}
This work is financially supported by National Center for Cognitive Research of ITMO University. This work was supported by the Ministry of Science and Higher Education of Russian Federation, research project no. 075-03-2020-139/2 (goszadanie no. 2019-1339).

This research is also supported supported by Enterprise Singapore ``Startup SG Tech POV'' Grant Scheme under SOMIN PTE LTD project.

\bibliographystyle{ACM-Reference-Format}
\bibliography{ref}

\end{document}